\documentclass[aip, amsmath,amssymb,preprint,numerical]{revtex4-2}

\usepackage{graphicx}               
\usepackage{wasysym}
\usepackage{lmodern}                
\usepackage[T1]{fontenc}            
\usepackage[utf8]{inputenc}         
\usepackage{indentfirst}            
\usepackage{physics}
\usepackage{subcaption}
\usepackage{csquotes}
\usepackage{cleveref}
\usepackage{natbib}
\usepackage{float}
\usepackage{xcolor}
\begin{document}

\title{The effect of parameter drift in the transport of magnetized plasma particles}

\author{P. Haerter}%
 \email{haerter@ufpr.br}
\affiliation{Departamento de Física, Universidade Federal do Paraná, Curitiba, 81531-990, Paraná, Brazil.}
\author{R. L. Viana}%
\affiliation{Departamento de Física, Universidade Federal do Paraná, Curitiba, 81531-990, Paraná, Brazil.}
\affiliation{Centro Interdisciplinar de Ciência, Tecnologia e Inovação, Núcleo de Modelagem e Computação Científica, Universidade Federal do Paraná, Curitiba, 81531-990, Paraná, Brazil}

\date{\today}

\begin{abstract}
We investigate how time dependent modulations of drift wave amplitudes affect particle transport and chaos in a magnetized plasma. Using the Horton model, we apply a sawtooth ramp to a primary wave's amplitude and periodic rectangular kicks to secondary waves, simulating a driven system. Particle transport is quantified by the Mean Square Displacement (MSD) exponent, $\alpha$, and chaos by the Maximum Lyapunov Exponent (MLE). Our primary finding is a strong negative correlation between the system's average chaoticity and its transport efficiency. We show that rapid sawtooth ramping (short period $\tau$) produces highly efficient, superdiffusive transport ($\alpha > 1$). In contrast, slower ramping increases the system's chaos but suppresses transport, driving it towards normal diffusion ($\alpha \to 1$). This counter intuitive result demonstrates that heightened chaos destroys the coherent, streamer like structures necessary for superdiffusive flights. Our findings indicate that the coherence of the turbulent field, rather than its raw chaoticity, is the key determinant of transport efficiency, offering a new perspective on plasma control.

\end{abstract}
\maketitle

\section{Introduction}

Achieving sustained nuclear fusion in magnetic confinement devices like tokamaks requires overcoming the challenge of anomalous transport, where turbulent fluctuations drive particles and energy out of the plasma core much faster than classical theories predict \cite{carrerasProgressAnomalousTransport1997,hortonTurbulentTransportMagnetized2017,hortonDriftWaveTurbulence2008,balescuAspectsAnomalousTransport2005,chenIntroductionPlasmaPhysics2016,garbetPhysicsTransportTokamaks2004}. This turbulence is often driven by low frequency electrostatic drift waves at the plasma edge \cite{hortonDriftWavesTransport1999}, which generate stochastic electric fields and induce chaotic particle motion via the $\mathbf{E} \times \mathbf{B}$ drift \cite{camargoInfluenceMagneticFluctuations1996,hasegawaPseudothreedimensionalTurbulenceMagnetized1978}. Cyclical magnetohydrodynamic (MHD) events in the plasma core, such as the sawtooth instability, can act as a natural pump, periodically expelling energy that modulates the conditions driving this edge turbulence \cite{zhaoEffectsSawtoothHeat2022,wessonSawtoothOscillations1986}.

Historically, plasma turbulence was often viewed as a source of purely random, diffusive transport. However, a more modern understanding reveals that turbulence can self organize into coherent structures that profoundly regulate transport \citep{diamondZonalFlowsPlasma2005,balescuAspectsAnomalousTransport2005,terrySuppressionTurbulenceTransport2000}. These structures include zonal flows, poloidally symmetric sheared flows that can act as transport barriers, and streamers, which are radially elongated convective cells that create rapid transport highways \citep{xuBlobHoleFormation2009,scottEnergeticsInteractionElectromagnetic2005}. The overall confinement efficiency is therefore determined by the complex competition between the randomizing effect of chaos and the formation of these ordered, coherent structures.

In this work, we investigate how the temporal dynamics of drift wave amplitudes, motivated by these cyclical plasma phenomena, govern the balance between chaotic and coherent transport. We employ the Horton drift wave model\cite{hortonOnsetStochasticityDiffusion1985} as our framework but introduce a crucial extension: we impose time dependent modulations on the wave amplitudes to mimic the energy dynamics of a driven dissipative system . Specifically, the primary wave's amplitude is modulated by a sawtooth ramp, representing a slow, periodic energy pump. The secondary waves are modulated by periodic rectangular pulses, modeling fast, intermittent energy transfer between modes, a scenario analogous to nonlinear three wave interactions \citep{haerterChaoticTransportDrift2025,batistaNonlinearThreemodeInteraction2006}. This approach allows us to explore how the interplay between slow pumping and fast kicks dictates the resulting transport regime.

Our analysis, based on parameters relevant to experimental devices like the TCABR Tokamak \citep{nascimentoPlasmaConfinementUsing2005,ferreiraTurbulenceTransportScrapeoff2004,castroTemperatureFluctuationsPlasma1996}, reveals two distinct transport regimes. We find that rapid sawtooth ramping (short period $\tau$) promotes the formation of coherent, streamer like transport channels, leading to highly efficient, superdiffusive motion ($\alpha > 1$). In contrast, slower ramping (long period $\tau$) increases the chaoticity of the wave field. This heightened chaos disrupts the streamers, breaking the long range correlation of particle trajectories and driving the system towards less efficient, normal diffusion ($\alpha \to 1$). Our central finding is therefore a strong negative correlation between the degree of chaos and the efficiency of anomalous transport, underscoring the critical role of wave field coherence in plasma confinement \citep{haerterChaoticTransportDrift2025}.

This paper is structured as follows: Section~\ref{sec:drift_waves_model} details the time dependent drift wave model. Section~\ref{sec:particle_transport} presents the analysis of particle transport via the Mean Square Displacement. Section~\ref{sec:MLE} provides the chaos analysis using the Maximum Lyapunov Exponent and connects it to the observed transport regimes. Finally, Section~\ref{sec:conclusions} summarizes our findings and their implications for controlling transport in magnetized plasmas.

\section{Drift Waves model}
\label{sec:drift_waves_model}

\begin{figure}[ht!]
    \centering
	{\includegraphics[scale=0.12]{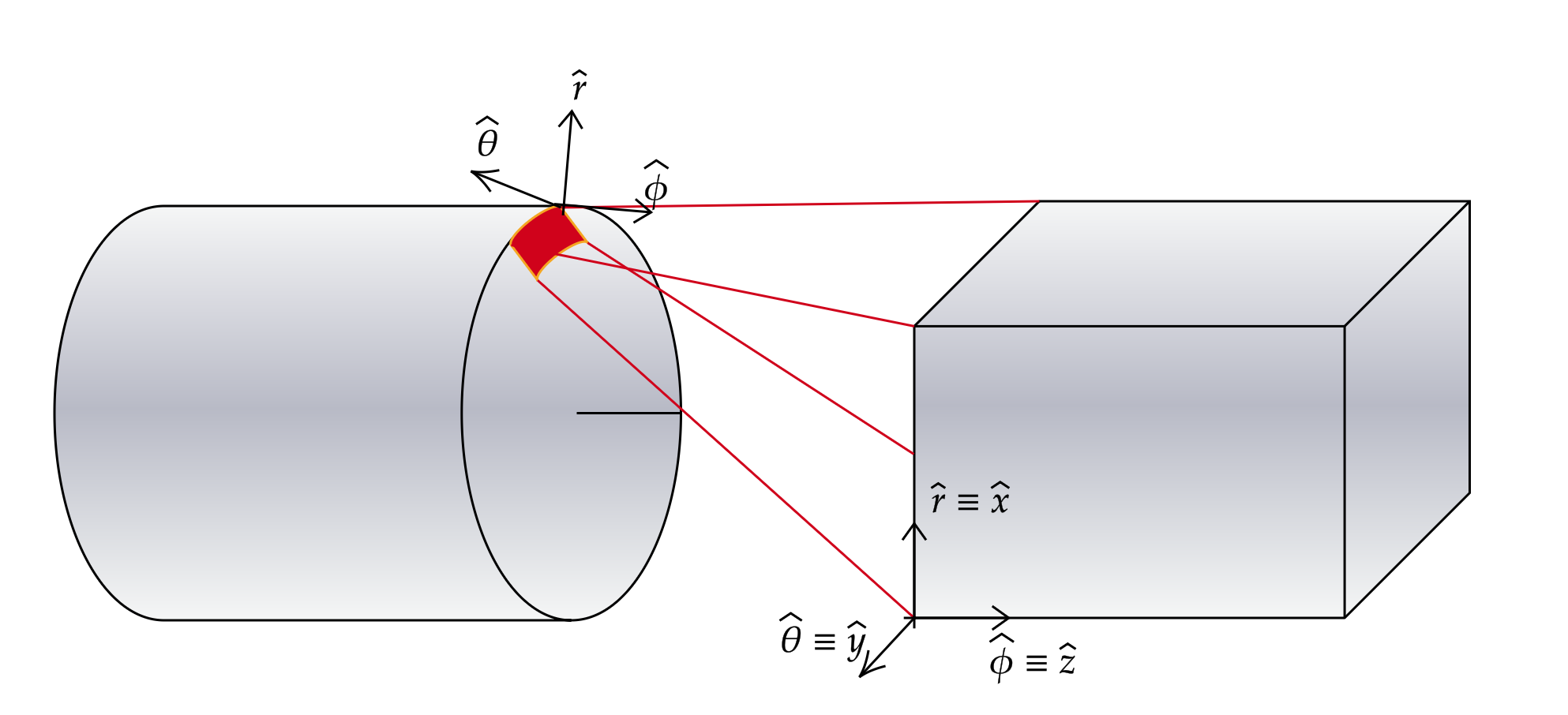}}
    \caption{Graphical representation of the transformation of cylindrical coordinates to rectangular ones.}
    \label{fig:Diagrama}
\end{figure}

The dynamics of particles subjected to sawtooth wave pumping can be described by multiple models. For particles at the plasma edge, the model proposed by Horton is particularly suitable\cite{hortonOnsetStochasticityDiffusion1985}. In this region, by neglecting curvature effects, the toroidal geometry can be locally approximated by a Cartesian slab. We therefore transform the toroidal coordinates $(r, \theta, \phi)$ into a Cartesian system $(x, y, z)$, where $x$ represents the radial, $y$ the poloidal, and $z$ the toroidal directions, as illustrated in Fig.~\ref{fig:Diagrama}. The uniform magnetic field is directed along the z axis, given by $\vec{B} = B_0\hat{z}$.

We assume that the ion and electron charges do not appreciably affect the electric and magnetic fields; hence, the corresponding guiding centers can be considered as passive tracers, or test particles. Such test particles move with the $\vb{E} \cross \vb{B}$ drift velocity.

\begin{equation}
    \vb{v} = - \frac{\nabla \Phi \cross \vb{B}}{B^2},
\end{equation}
where the electric field is written as $\vb{E} = -\nabla \Phi$, with $\Phi(x,y,t)$ being a scalar potential. This system is equivalent to the following set of canonical equations:
\begin{align}
    v_x &= \dv{x}{t} = -\frac{1}{B_0}\pdv{\Phi}{y}, \label{eq:vx} \\
    v_y &= \dv{y}{t} = \frac{1}{B_0}\pdv{\Phi}{x}, \label{eq:vy}
\end{align}
corresponding to the time dependent Hamiltonian
\begin{equation}
    H(x,y,t) = \frac{1}{B_0}\Phi(x,y,t).
\end{equation}

The electric potential can be divided into two parts: an equilibrium part $\Phi_0(x)$ corresponding to a radial electric field, and a perturbation caused by $N$ drift waves with amplitudes $A_i$, frequencies $\omega_i$, and wave vectors $\vb{k}_i = (k_{x,i}, k_{y,i})$ for $i = 1, 2, \ldots, N$. This gives:
\begin{equation}
    \Phi(x, y, t) = \Phi_0(x) + \sum_{i=1}^{N} A_i \sin(k_{x,i} x) \cos(k_{y,i} y - \omega_i t),
    \label{eq:Horton-phi}
\end{equation}
where we assume a stationary wave pattern along the radial direction $x$ and a traveling wave along the poloidal direction $y$. Let $L_x$ and $L_y$ be the characteristic lengths along these directions, such that:
\begin{equation}
    k_x = \frac{n \pi}{L_x}, \qquad k_y = \frac{2 \pi m}{L_y},
\end{equation}
where $m$ and $n$ are suitably chosen positive integers.

In the case of $N$ drift waves, the Hamiltonian reads:
\begin{equation}
    H(x,y,t) = \frac{1}{B_0}[\Phi_0(x) + \sum_{i=1}^{N} A_i \sin(k_{x,i} x) \cos(k_{y,i} y - \omega_i t)].
\end{equation}

After transitioning to a reference frame moving with the phase velocity of the first wave, $u_1$, and considering the resonant case where the drift velocity satisfies 
$$
v_e = \frac{1}{B}\dv{\Phi_0}{x} = u_1 = \frac{\omega_1}{k_{y,1}},
$$
the Hamiltonian becomes:
\begin{eqnarray}
    H(x,y,t) = A_1 \sin(k_{x,1} x)\cos(k_{y,1} y) +\nonumber \\ \sum_{i=2}^{N} A_i \sin(k_{x,i} x + \beta_i) \cos\big(k_{y,i}(y - u_i t)\big),
    \label{eq:H_Final}
\end{eqnarray}

where $u_i = \frac{\omega_i}{k_{y,i}} - \frac{\omega_1}{k_{y,1}}$ and $\beta_i$ is a phase term associated with the breakdown of transport barriers in symmetry regions. This phase enables particle transport along the $x$-direction, which is the focus of this study\cite{klevaStochasticExBParticle1984}.

\begin{figure*}[ht!]
    \centering
	\includegraphics[scale=0.5]{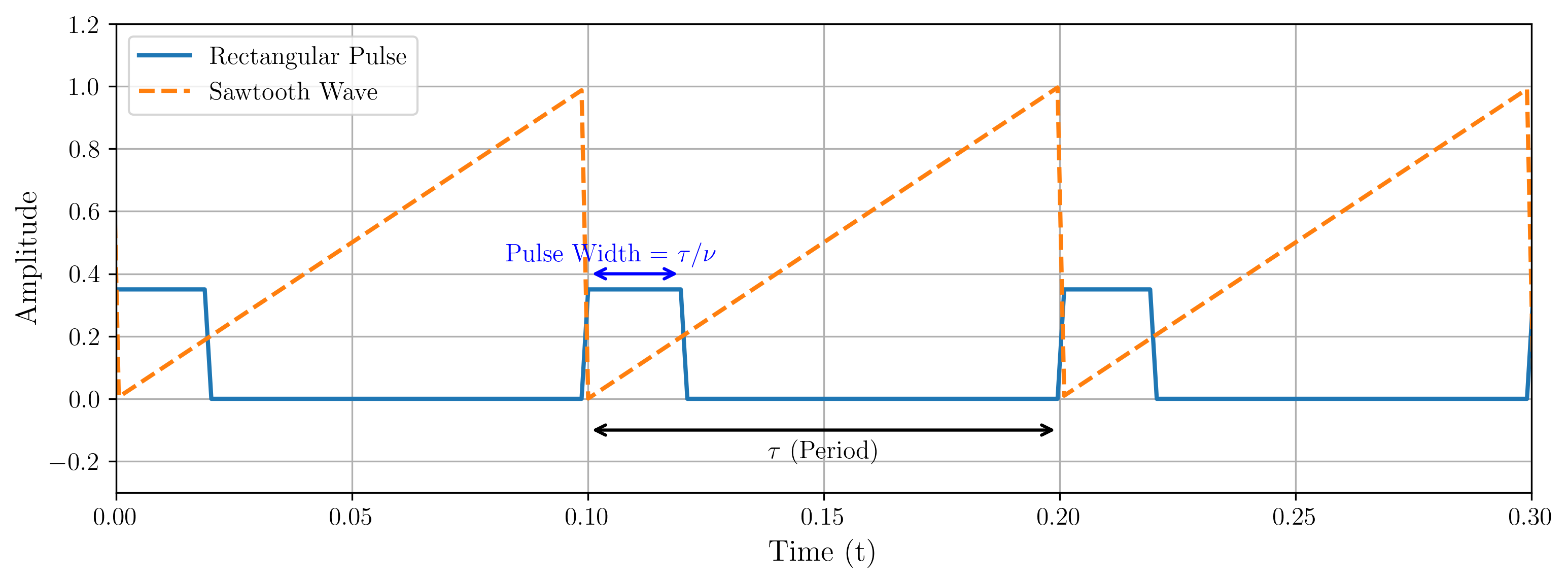}
    \caption{Time evolution of the modulated wave amplitudes. The dashed orange line shows the sawtooth modulation of the primary wave amplitude, $A_1(t)$, with period $\tau$. The solid blue line shows the rectangular pulse modulation for the secondary wave amplitudes, $A_{i>1}(t)$, which represents periodic kicks to the system.}
    \label{fig:waveforms}
    \label{fig:Serra_Graf}
\end{figure*}

The sawtooth pumping\cite{wessonSawtoothOscillations1986} is introduced by modulating the amplitude of the reference wave, $A_1$, as follows:
\begin{equation}
    A_1(t) = A_{1,max} \frac{t \pmod \tau}{\tau}
\end{equation}
where $A_{1,max}$ is the maximum amplitude (set to 1.0 in this work) and $\tau$ is the period of the sawtooth ramp.

Simultaneously, the amplitudes of the other waves ($A_i$ for $i > 1$) are modulated as periodic rectangular pulses to represent energy exchange between modes\cite{batistaNonlinearThreemodeInteraction2006}. Their amplitude is given by:
\begin{equation}
    A_{i>1}(t) = 
    \begin{cases} 
      A_{i,max} & \text{if } t \pmod \tau \le \frac{\tau}{\nu} \\
      0 & \text{otherwise}
    \end{cases}
\end{equation}
where $A_{i,max}$ is the maximum amplitude of the pulse, $\tau$ is the period, and the parameter $\nu$ controls the pulse width, $W = \tau/\nu$.

Considering three modes\cite{katouResonantThreeWaveInteraction1982} in the Hamiltonian equation \eqref{eq:H_Final}, we evolved the canonical equations \eqref{eq:vx} and \eqref{eq:vy} numerically from $t_i = 0$ to $t_f = 10^4$. A Poincaré map was constructed by sampling $(x,y)$ values at times equal to integer multiples of the period $T = 2\pi/\omega_1$. The system parameters are defined as follows:
\begin{align*}
    k_{x,1} &= 5,      & k_{y,1} &= 6,    & \omega_1 &= 2,   & A_1 &= 1.0, \\
    k_{x,2} &= -3,     & k_{y,2} &= -3,   & \omega_2 &= 2,   & A_2 &= 0.1, \\
    k_{x,3} &= -2,     & k_{y,3} &= -3,   & \omega_3 &= 2,   & A_3 &= 0.033.
\end{align*}
These values were chosen to reflect realistic plasma wave dynamics\cite{hellerCorrelationPlasmaEdge1997,batistaNonlinearThreemodeInteraction2006,castroTemperatureFluctuationsPlasma1996,dossantoslimaBicoherenceElectrostaticTurbulence2009}, with $\tau$ (the ramping period) and $\nu$ (the duration of the kicks) serving as a tunable parameter. The representation of the waves amplitudes can be seen in Fig.~\ref{fig:Serra_Graf}

\section{Particle Transport}
\label{sec:particle_transport}

To analyze the transport dynamics induced by the different combinations of ramping and pulse parameters, we simulated an ensemble of passive particles initially distributed uniformly in the domain $x \in [0, \pi]$ and $y \in [0, 2\pi]$. We employed the mean square displacement (MSD) to provide a global characterization of the chaotic transport by quantifying the spatial spread of particles over time.

\begin{figure*}[ht!]
    \centering
    \includegraphics[width=0.8\textwidth]{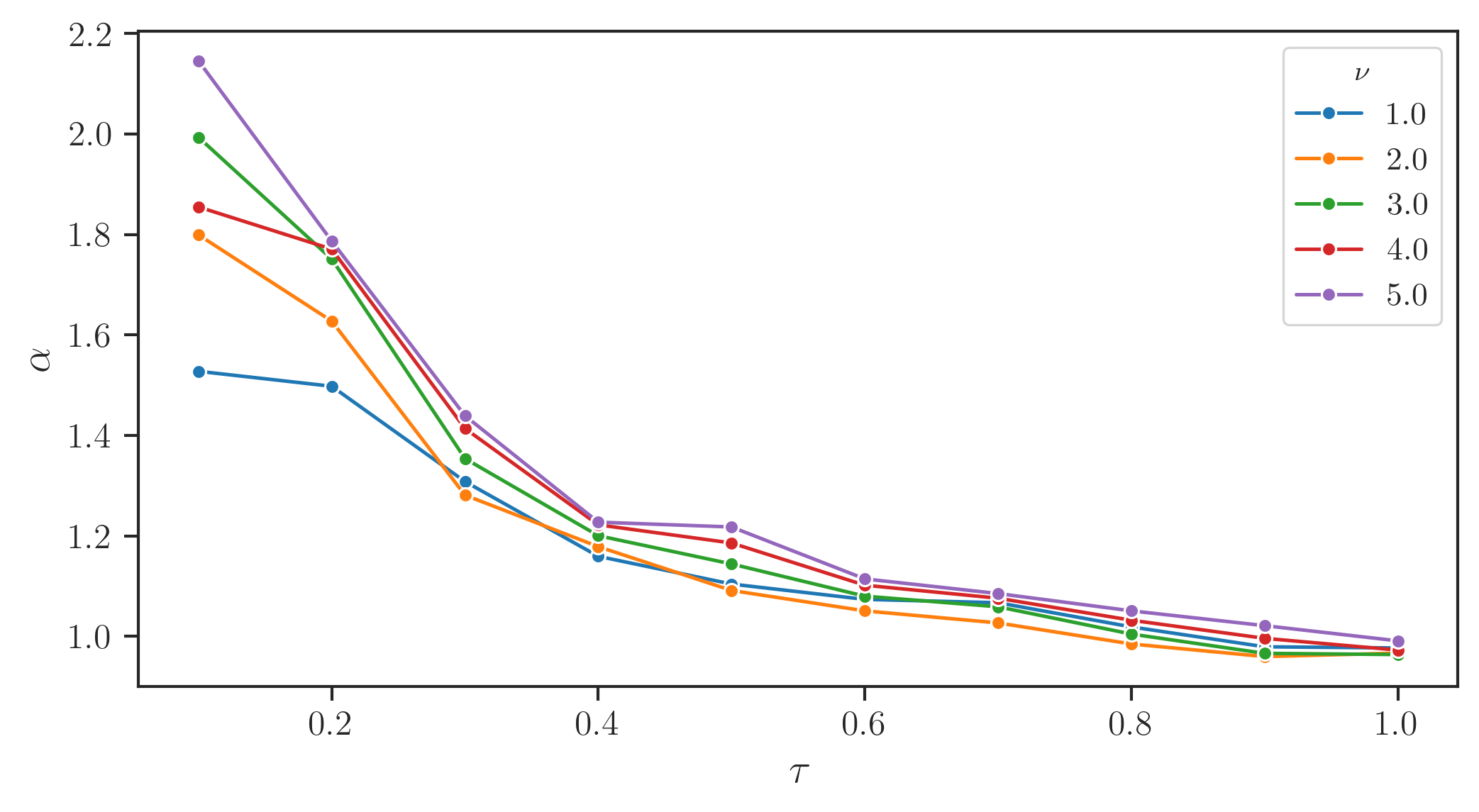}
    \caption{The MSD scaling exponent $\alpha$ as a function of the sawtooth wave period ($\tau$) for different values of the kick duration parameter ($\nu$). The plot demonstrates a clear suppression of anomalous transport (i.e., $\alpha$ approaches 1) as the ramping period $\tau$ is increased.}
    \label{fig:MSD}
\end{figure*}
\subsection{Mean Square Displacement}

For a statistical characterization of transport, we computed the mean square displacement (MSD) in the radial direction:
\begin{equation}
    \langle\sigma_x(t)^{2}\rangle = \frac{1}{N}\sum_{i=1}^{N}|x_{i}(t)-x_{i}(0)|^{2},
\end{equation}
where $N$ is the total number of particles. The long term behavior of the MSD scales with time as $\langle\sigma_x(t)^{2}\rangle \propto t^{\alpha}$, where the scaling exponent $\alpha$ classifies the transport regime. The regimes are subdiffusive ($\alpha < 1$), normal diffusion ($\alpha = 1$), superdiffusive ($1 < \alpha < 2$), or ballistic ($\alpha = 2$). An exponent of $\alpha > 2$ is sometimes referred to as hyperdiffusive.

Our analysis, summarized in Fig.~\ref{fig:MSD}, reveals a strong dependence of the transport regime on the sawtooth period $\tau$. For short ramping periods (small $\tau$), the particles exhibit highly enhanced, superdiffusive transport, with scaling exponents $\alpha$ approaching 2.0. As the ramping period $\tau$ increases, the transport is consistently suppressed, and the scaling exponent $\alpha$ trends towards 1.0, indicative of normal diffusion. This demonstrates that the system's transport properties can be effectively tuned from a superdiffusive or even hyperdiffusive state down to a standard diffusive process simply by adjusting the ramping frequency of the primary wave.

The parameter $\nu$, which controls the duration of the rectangular pulse kicks ($W = \tau/\nu$), also has a notable effect on the transport. Figure~\ref{fig:MSD} shows that shorter kicks (corresponding to larger $\nu$) generally lead to slightly higher values of the transport exponent $\alpha$. While shorter, the kicks still inject energy into the system, and their interaction with the primary ramping wave appears crucial for generating the most efficient transport.

Interestingly, this combination is a necessary mechanism for the emergence of the hyperballistic regime ($\alpha \ge 2$). It has been shown in previous work~\cite{haerterEscapeTransportChaotic2025} that a simpler model—one with a constant amplitude primary wave and suppressed secondary waves—only produces standard superdiffusive transport ($1 < \alpha < 2$). Therefore, the interplay between the slow sawtooth ramp of the main wave and the fast, periodic kicks from the other waves is the key physical process that gives rise to the hyperballistic motion observed in our results.

The observed transport regimes can be understood by considering the underlying phase space structure. The highly superdiffusive transport at small $\tau$ suggests the presence of coherent structures that allow for long, correlated particle flights. The rapid sawtooth ramp likely accelerates particles along preferential pathways before they can be fully randomized. In contrast, as the ramping period $\tau$ increases, the slower but more sustained perturbations lead to the widespread destruction of these coherent pathways. This enhanced chaos, which we will quantify in the next section, disrupts the long range correlations, causing particles to undergo a more random, diffusive motion and thus suppressing the overall transport efficiency ($\alpha \to 1$).

\section{Maximum Lyapunov Exponent}
\label{sec:MLE}
The time dependent Hamiltonian system described by Eq.~\ref{eq:H_Final} is non integrable, and as such, it exhibits chaotic dynamics for certain parameter ranges\cite{mathiasFractalStructuresChaotic2017}. The degree of this chaos can be quantified by measuring the average rate of separation of initially close trajectories in phase space. This rate is given by the Maximum Lyapunov Exponent (MLE), denoted by $\lambda$. For a given trajectory, the MLE is defined as:
\begin{equation}
    \lambda = \lim_{t \to \infty} \frac{1}{t} \ln \frac{\|\delta\mathbf{x}(t)\|}{\|\delta\mathbf{x}(0)\|},
	\label{eq:MLE}
\end{equation}
where $\|\delta\mathbf{x}(t)\|$ is the magnitude of the separation vector between two initially close trajectories at time $t$\cite{lichtenbergRegularChaoticDynamics1992,ottChaosDynamicalSystems2002}. A positive MLE ($\lambda > 0$) is a hallmark of chaos, indicating exponential divergence of trajectories and sensitive dependence on initial conditions. Conversely, a value of $\lambda \le 0$ signifies regular, non chaotic motion.

\begin{figure*}
\centering
\includegraphics[width=0.9\linewidth]{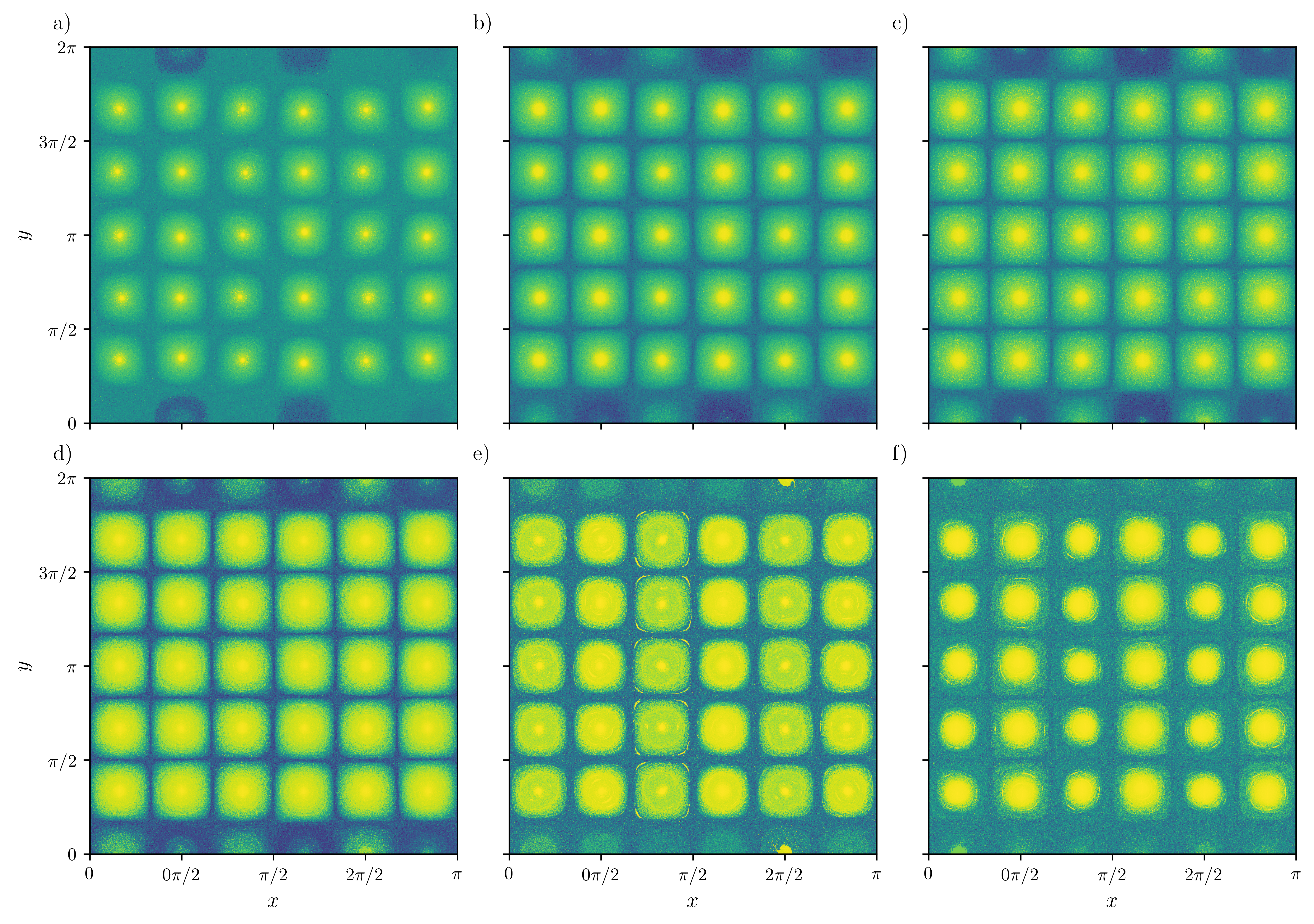}
\caption{
    Colormaps of the Maximum Lyapunov Exponent ($\lambda$) in the system phase space for fixed $\tau=0.1$ and varying $\nu$ in the top row: (a) $\nu=1.0$, (b) $\nu=3.0$, (c) $\nu=5.0$. The bottom row shows results for fixed $\nu=5.0$ and varying $\tau$: (d) $\tau=0.2$, (e) $\tau=0.4$, (f) $\tau=0.6$. The viridis colormap is used, where bright yellow corresponds to regular motion ($\lambda \approx 0$) and dark blue/purple indicates highly chaotic regions ($\lambda > 0$).    }
\label{fig:MLyapunov}
\end{figure*}

To visualize the structure of the phase space, we computed the MLE for a grid of initial conditions, as shown in the colormaps of Fig.\ref{fig:MLyapunov}. In these plots, regions of regular, quasi periodic motion (e.g., stable islands) correspond to low $\lambda$ values (bright yellow), while the chaotic sea is characterized by high $\lambda$ values (dark blue/purple).

The plots reveal clear trends with respect to the control parameters. As the sawtooth period $\tau$ is increased (Fig.\ref{fig:MLyapunov}d-f), the dark, chaotic sea expands and intensifies, encroaching upon the stable yellow islands. Conversely, as the kick duration is made shorter by increasing $\nu$ (Fig.\ref{fig:MLyapunov}a-c), the yellow regions of regular motion become more prominent, indicating a less chaotic system overall.

\begin{figure}
\centering
\includegraphics[width=0.9\linewidth]{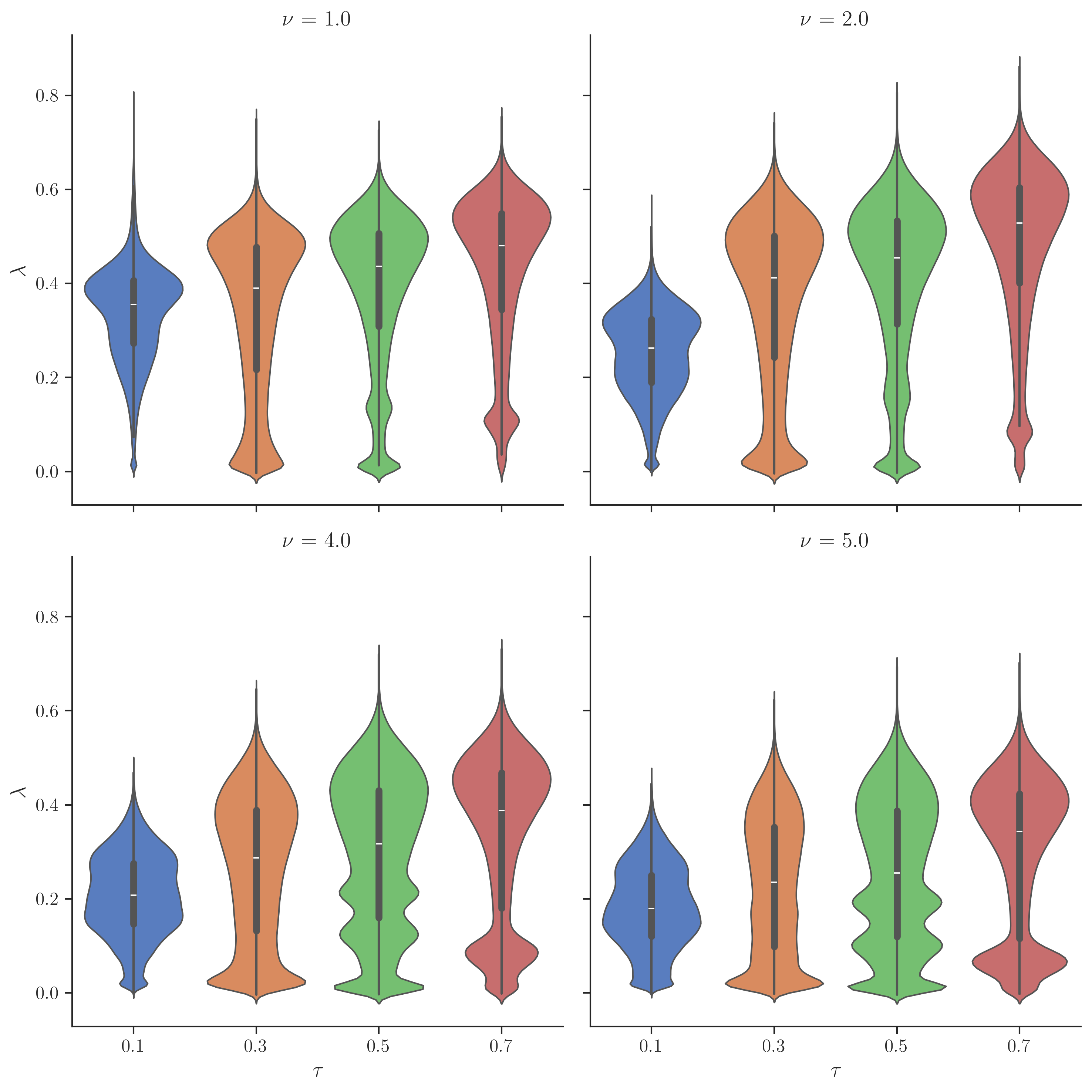}
\caption{
    Violin plots of the MLE ($\lambda$) distributions. Each panel corresponds to a fixed value of $\nu$. Within each panel, the distributions for different sawtooth periods $\tau$ are shown. The plots illustrate that chaos generally increases with $\tau$ and decreases with $\nu$.    }
\label{fig:MLE_Dist}
\end{figure}

\begin{figure}
\centering
\includegraphics[width=0.9\linewidth]{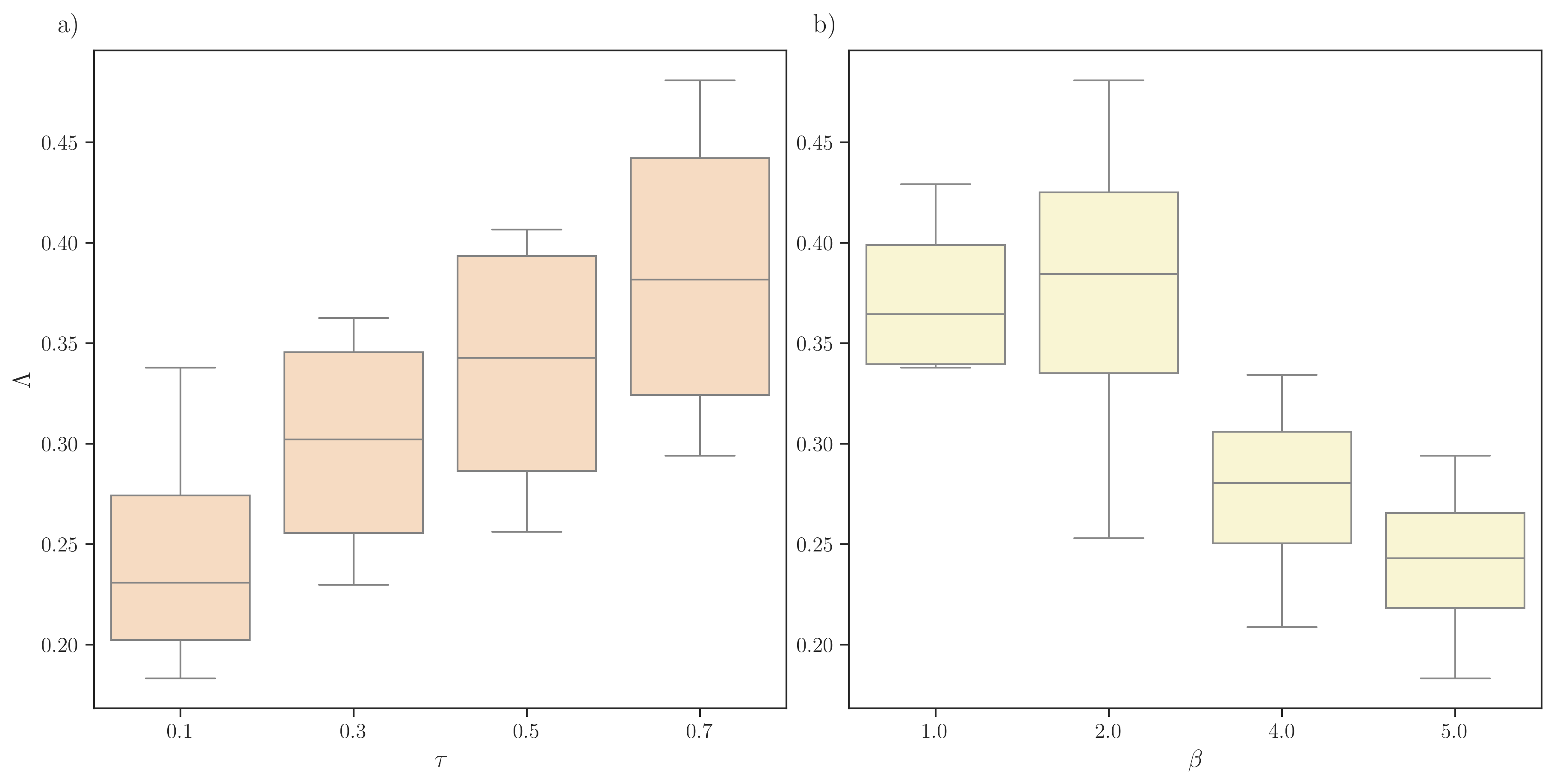}
\caption{
    Boxplots of the mean Lyapunov exponent, $\Lambda$. (a) $\Lambda$ as a function of the ramping period $\tau$. (b) $\Lambda$ as a function of the kick parameter $\nu$. The plots confirm that the average chaoticity of the system increases with $\tau$ and decreases with $\nu$.    }
\label{fig:BoxPlot_MMLE}
\end{figure}

A statistical overview of these trends is presented in the violin plots of Fig.~\ref{fig:MLE_Dist}, which show the distribution of MLE values for an ensemble of particles. The distributions confirm that increasing $\tau$ shifts the weight of the distribution towards higher $\lambda$ values (more chaos), while increasing $\nu$ shifts it towards zero (less chaos). To summarize this behavior, we compute the mean MLE over the entire ensemble, which we denote as $\Lambda$. Fig.~\ref{fig:BoxPlot_MMLE} shows boxplots of $\Lambda$ versus $\tau$ and $\nu$, reinforcing these conclusions.

\begin{figure}
\centering
\includegraphics[width=0.9\linewidth]{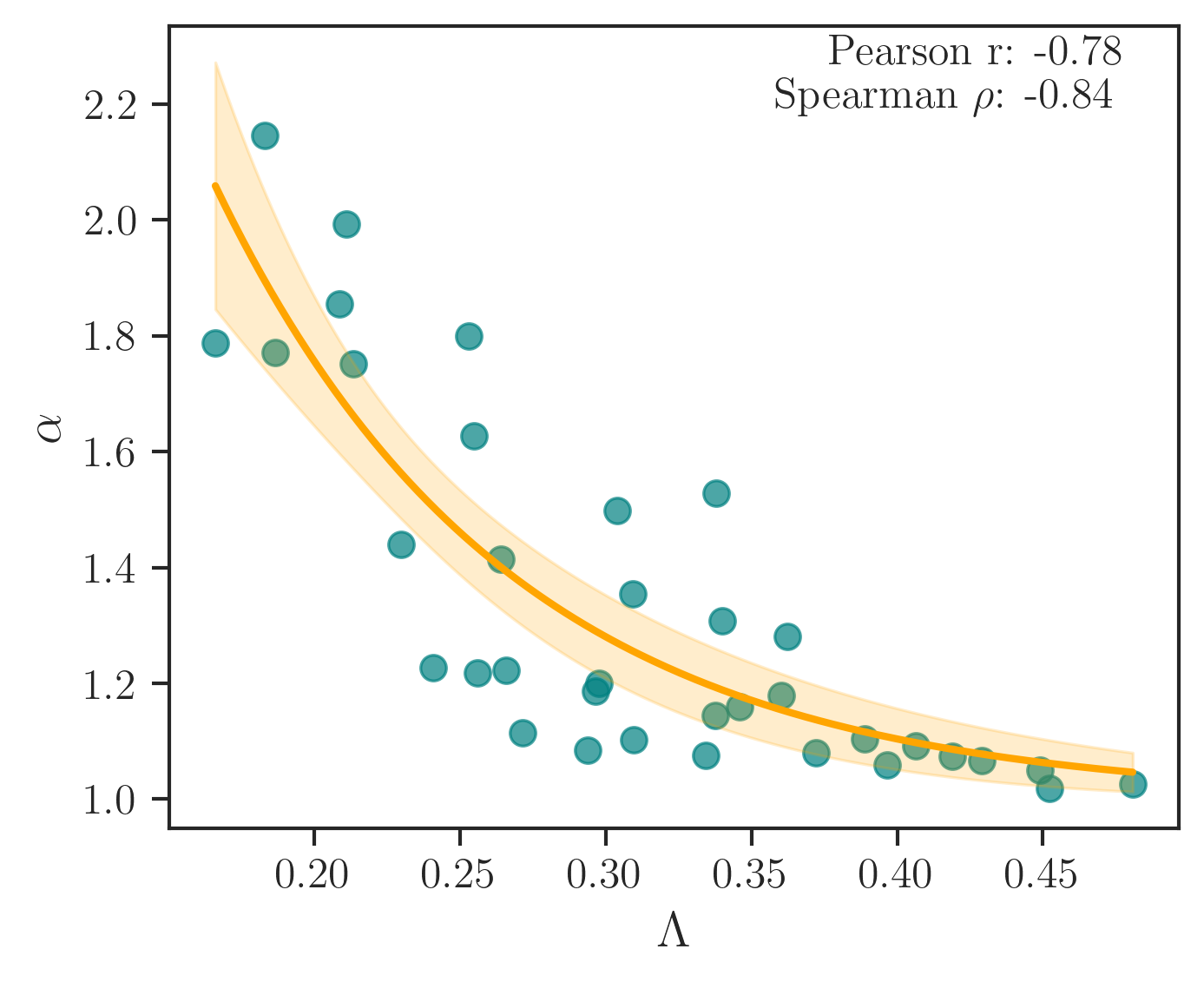}
\caption{
     Scatter plot showing the relationship between the mean Lyapunov exponent ($\Lambda$) and the transport exponent ($\alpha$). The strong negative correlation indicates that more chaotic systems exhibit less anomalous transport (i.e., $\alpha$ is closer to 1). The orange line is a best fit curve, with the shaded region representing the 95\% confidence interval.    }
\label{fig:RefPlot}
\end{figure}

The central result of this paper is the relationship between the system average chaoticity, $\Lambda$, and the transport exponent, $\alpha$. As shown in the scatter plot in Fig.~\ref{fig:RefPlot} , there is a strong negative correlation between these two quantities (Pearson r = -0.78, Spearman $\rho$ = -0.84). This indicates that more chaotic systems produce less efficient, more diffusive transport.

This seemingly counter intuitive result can be explained by the disruption of coherent structures in phase space. For systems with low chaos (low $\tau$, high $\nu$), particles can engage in long, correlated flights along stable manifolds, leading to superdiffusive motion (high $\alpha$). However, as chaos increases (high $\tau$, low $\nu$), these coherent pathways are destroyed. Particles are scattered more effectively throughout the chaotic sea, their trajectories become randomized, and their motion resembles a classic random walk. This destruction of long range correlation suppresses the transport efficiency, driving the system towards a state of normal diffusion ($\alpha \to 1$).

\section{Conclusions}
\label{sec:conclusions}

In this work, we investigated the impact of time dependent wave amplitude modulations on particle transport and chaos in a magnetized plasma. By extending the Horton drift wave model to include a primary wave driven by a slow sawtooth ramp and secondary waves subjected to fast, periodic kicks, we simulated a dynamic environment analogous to a driven dissipative system. Using the Mean Square Displacement (MSD) and the Maximum Lyapunov Exponent (MLE) as our primary diagnostics, we systematically explored the effects of the ramping period ($\tau$) and the kick duration parameter ($\nu$).

Our central finding is that the transport regime is highly tunable and exhibits a strong, counter intuitive negative correlation with the system's chaoticity. We demonstrated that rapid sawtooth ramping (small $\tau$) facilitates highly efficient, superdiffusive transport ($\alpha > 1$), which can even enter a hyperballistic regime. Conversely, slower ramping (large $\tau$) increases the overall chaos in the system but suppresses anomalous transport, driving the scaling exponent towards that of normal diffusion ($\alpha \to 1$).

This result is explained by the competition between chaos and the formation of coherent transport structures. In less chaotic regimes, particles can follow streamer like channels, leading to long, correlated flights and superdiffusive behavior. As the system becomes more chaotic, these coherent pathways are destroyed. The increased scattering randomizes particle trajectories, disrupting the mechanism for efficient transport and resulting in a classic diffusive process. This highlights a crucial insight: simply increasing the chaoticity of the driving fields does not necessarily enhance transport; instead, it can suppress it by destroying the underlying coherence that enables it.

The findings have significant implications for controlling plasma transport. They suggest that the nature of particle flux is not a fixed outcome of turbulence but can be actively manipulated by controlling the temporal characteristics of the driving instabilities. Modulating the frequency and form of energy injection into turbulent modes could offer a potential strategy to either enhance the flushing of impurities or suppress the transport of fuel particles to improve confinement. Future work could explore more complex and realistic modulation waveforms derived from advanced simulations or experimental data, as well as incorporate the effects of magnetic field fluctuations to build a more comprehensive picture of transport regulation in magnetized plasmas.

\begin{acknowledgments}
This work has been supported by grants from the Brazilian Government Agencies CNPq and CAPES. P. Haerter received partial financial support from the following Brazilian government agencies: CNPq (140920/2022-6), CAPES (88887.898818/2023-00).  R. L. Viana received partial financial support from the following Brazilian government agencies: CNPq (403120/2021-7, 301019/2019-3), CAPES (88881.143103/2017-01). 
\end{acknowledgments}

\section*{AUTHOR DECLARATIONS}

\subsection*{Conflict of Interest}
The authors have no conflicts to disclose.

\subsection*{DATA AVAILABILITY}
Data sharing is not applicable to this article as no new data were created or analyzed in this study.

\section*{References}
\bibliography{Refs}

\end{document}